\documentclass{PoS}

\usepackage{amsmath,amssymb}
\usepackage{graphicx,psfrag,here,epsfig}
\usepackage{longtable,multirow}

\newcommand{\fs}{\; .}

\def\co{\; ,}

\newcommand{\nn}{\nonumber\\[1.5ex]}
\newcommand{\ed}{\end{document}}

\newcommand{\myfrac}[2]{\mbox{\large{$\frac{#1}{#2}$}}}

\newcommand{\mk}{\bar{M}_K}

\renewcommand{\L}{\mathcal{L}}
\newcommand{\osn}[1]{\oldstylenums{#1}}
\newcommand{\qcd}{\sc qcd\rm}
\newcommand{\eom}{\sc eom\rm}
\newcommand{\chpt}[1]{$\chi\text{\sc pt\rm}_{#1}$}
\newcommand{\lecs}{\sc lec\rm s}
\newcommand{\lec}{\sc lec\rm }
\newcommand{\SU}[1]{\mathrm{SU}(#1)}
\renewcommand{\d}{\partial}
\newcommand{\cO}{\mathcal{O}}

\newcommand{\clr}[1]{c^{\mathrm{r}}_{#1}}
\newcommand{\Clr}[1]{C^{\mathrm{r}}_{#1}}
\newcommand{\TClr}{2\Clr{13}-\Clr{11}}

\newcommand{\lo}{\sc lo\rm}
\newcommand{\nlo}{\sc nlo\rm}
\newcommand{\nnlo}{\sc nnlo\rm}
\title{Relations between SU(2)- and SU(3)-LECs in chiral perturbation theory}
\ShortTitle{Matching {\footnotesize LEC}s in $\chi${\footnotesize PT}}

\author{Juerg Gasser $^a$, 
Christoph Haefeli $^a$,
\speaker{Mikhail A. Ivanov}$^{\,\,b}$,
       Martin Schmid $^a$ \\

$^a$Albert Einstein Center for Fundamental Physics,
     Institute for Theoretical Physics,\\ University of Bern,
    Sidlerstr. 5, CH--3012 Bern, Switzerland 
         \thanks{
 The Albert Einstein Center for Fundamental Physics is
  supported by the ``Innovations- und Kooperationsprojekt C-13'' of
  the ``Schweizerische Universit\"atskonferenz SUK/CRUS''.
This work was  in addition supported  by the Swiss
National Science Foundation,   by EU MRTN-CT-2006-035482
(FLAVIAnet)
and by the Helmholtz Association through funds provided to the 
virtual institute 
``Spin and strong QCD'' (VH-VI-231).
}

$^b$ Bogoliubov  Laboratory of Theoretical Physics,
Joint Institute for Nuclear Research, \\
141980 Dubna (Moscow region), Russia \\

        E-mail: \email{gasser@itp.unibe.ch},
                \email{haefeli@itp.unibe.ch},\newline
\hspace*{1.1cm} \email{ivanovm@theor.jinr.ru},
                \email{schmidm@itp.unibe.ch}
}

\abstract{
Chiral perturbation theory in the two--flavour sector 
allows one to analyse Green functions in \qcd{} in the limit where 
the strange quark mass is considered to be large in comparison to the 
external momenta and to the light quark masses $m_u$ and $m_d$.
In this framework, the low--energy constants of $\SU{2}_R\times \SU{2}_L$ 
depend  on the value of the heavy quark masses. 
For the coupling constants 
which occur at order $p^2$ and $p^4$ in the chiral expansion,
 we worked out in Ref.~\cite{LECs-p4} 
the dependence on the strange quark mass at two--loop accuracy, and provided 
 in Ref.~\cite{LECs-p6} analogous relations for some of the  couplings $c_i$ which 
are relevant at order  $p^6$.
 This talk comments on the methods used, and illustrates implications  
of the results obtained.
}

\FullConference{6th International Workshop on Chiral Dynamics\\
                 July 6-10, 2009\\
                 Bern, Switzerland}

\begin{document}


\section{Introduction}
At low energies and small quark masses, the Green functions of quark currents  
can be analysed in the framework of chiral perturbation theory (\chpt{})
\cite{weinberg,glann,glnpb}. The method allows one to work out 
the momentum and quark mass dependence of the quantities of interest in a 
systematic and coherent manner.
It is customary to perform the quark mass expansion either around
$m_u=m_d=0$, with the strange quark mass held fixed at its physical value
(\chpt{2}), or to consider an expansion in all three quark masses around
$m_u=m_d=m_s=0$ (\chpt{3}).
The corresponding  effective Lagrangians contain 
low--energy constants (\lecs{}) that parametrise the degrees of freedom 
which are integrated out. The two expansions are not
independent: in a particular limit specified below, \chpt{3} reduces to \chpt{2}. 
As a result of this,   
 one can express the \lecs{} in the two--flavour case through the ones in \chpt{3}, in a 
perturbative manner. The relations amount to a series expansion in 
the strange quark mass. Generically,
\begin{equation}\label{eq:matching}
k^\mathrm{r}=\sum_{m\geq -m_0} d_{m} z^m\,,\, z=\frac{m_sB_0}{(4\pi F_0)^2}\,.
\end{equation}
Here, $k^\mathrm{r}$  stands for any of the renormalized \chpt{2} \lecs{}, while 
$F_0,B_0$ are the \lecs{} at order $p^2$ in \chpt{3}. The coefficients $d_m$ (whose dependence on the chosen $k^\mathrm{r}$ is suppressed in the notation) contain  renormalized \lecs{} from \chpt{3}, and  powers of the logarithm $\ln{\frac{m_sB_0}{\mu^2}}$, 
where $\mu$ denotes the 
standard  renormalization scale. For  $k^\mathrm{r}$  at order $p^{2(N+1)}$, 
one has $m_0=N$, and the corresponding leading term $d_{m_0}z^{-m_0}$ is 
generated by tree graphs in \chpt{3}. The next-to-leading order term 
requires a one--loop calculation, etc. In the following, we refer to the relations 
(\ref{eq:matching}) as \emph{matching relations}, obtained by \emph{matching}  
\chpt{2} to \chpt{3}  in the specific limit mentioned. The matching relations are useful, 
because  they provide i) additional information on the \lecs{} in \chpt{3}, 
and ii) internal consistency checks.

For the \lecs{} at order $p^2$ and $p^4$, the matching was performed to one loop (to two 
loops) in  Ref. \cite{glnpb}  (Ref.~\cite{LECs-p4}), and for a subclass of  \lecs{} 
at order $p^6$ to two loops in Ref.~\cite{LECs-p6}.

We comment on related work which is available in the literature.
\renewcommand{\labelenumi}{\roman{enumi})}
\begin{enumerate}
\item The strange quark
mass expansion of the \chpt{2} \lec{} $B$ ($F^2B$) was provided at two--loop accuracy in Ref.~\cite{Kaiser:2006uv} (\cite{Mouss:Sigma}). 
\item Matching of the order $p^6$ \lecs{} in the parity--odd sector was performed recently in Ref.~\cite{kampf_moussallam}.
\item Analogous work was done in the baryon sector in 
Refs.~\cite{meissner_frink,Mai:2009ce}, and for electromagnetic interactions in
 Refs.~\cite{gasser_rusetsky,jallouli_sazdjian,NehmePhD,Haefeli:2007ey}. 
\end{enumerate}
\renewcommand{\labelenumi}{\arabic{enumi}.}

The outline of the talk is as follows. In Section~\ref{sec:example}, we
illustrate the matching for the pion vector form factor at
order $p^4$. In Section~\ref{sec:description}, we give a short description of
the method used to obtain the matching relations in general. In Section~\ref{sec:l2} we display the
structure of the results at order $p^2$ and $p^4$, and discuss in some detail the
matching relation for   $l_2^\mathrm{r}$ to illustrate its use, whereas Section~\ref{sec:results} concerns
the matching at order $p^6$. The final Section~\ref{sec:conclusion} 
contains concluding remarks. We refer the interested 
reader to Refs.~\cite{LECs-p4,LECs-p6} for more details, and for the full 
results of the matching relations.


\section{The pion vector form factor at order $\mathbf{p^4}$}
\label{sec:example}

We first illustrate how  the relations  between the \lecs{} emerge, and
 consider for this purpose the vector form factor of the pion,
 \begin{equation}
\langle \pi^+(p')\,|\tfrac{1}{2}(\bar u {\gamma_\mu} u-\bar d{\gamma_\mu}
d)|\pi^+(p)\rangle = (p+p')_\mu 
F_V(t)\,\,;\,\,
t=(p'-p)^2\,,   
 \end{equation}
in the chiral limit $m_u=m_d=0$. In the three--flavour case, at
one--loop order, the form factor  reads in $d$ space--time dimensions
\begin{equation}
  \label{eq:loopsu3}
F_{V,3}(t)
=1+\frac{t}{F_0^2}\left[\Phi(t,0;d)+\tfrac{1}{2}\Phi(t,M_K;d)\right] 
+ \frac{2L_9 t}{F_0^2}\,.  
\end{equation}
The loop function $\Phi$ is given by
\begin{equation}
\Phi(t,M;d)=\frac{\Gamma(2-\frac{d}{2})}{2(4\pi)^{d/2}}\int_0^1\!\mathrm{d}u\,
u^2\left[M^2-\myfrac{t}{4}(1-u^2)\right]^{\frac{d-4}{2}}\,.  
\end{equation}
It is generated by mesons of mass $M$ running in the loop [$M_K$ denotes 
the kaon mass at $m_u=m_d=0$, at order $p^2$].
 Furthermore,  $F_0$ stands for   the pion decay constant
at $m_u=m_d=m_s=0$, and $L_9$ is one of the \lecs{} in  \chpt{3} at order $p^4$. 

In \chpt{2}, the corresponding one--loop expression is
\begin{equation}
F_{V,2}(t)=1+\frac{t}{F^2} \Phi(t,0;d)-\frac{l_6t}{F^2}  \,\,,  
\end{equation}
where $F$ denotes the pion decay constant at $m_u=m_d=0,m_s\neq 0$, and where
$l_6$ is one of the  low--energy constants in \chpt{2} at order $p^4$. 
If one identifies $F$ with $F_0$ at this order, the  expressions $F_{V,3}$ and
$F_{V,2}$ still differ in the
coefficient of the term proportional to $t$, and in the contribution
$\Phi(t,M_K;d)$, which is absent in the two--flavour case, because kaons are
integrated out in that framework.

To proceed, we note that the loop function $\Phi(t,M;d)$ 
is holomorphic in the complex $t$--plane, cut
along the real axis for Re $t\geq 4M^2$. 
Therefore,$\Phi(t,0;d)$ develops a branch point at $t=0$, whereas $\Phi(t,M_K;d)$
reduces to a polynomial at  $t/M_K^2\ll 1$,
\begin{equation}
  \label{eq:polynom}
\Phi(t,M_K;d)=\sum_{l=0}^{\infty} 
\Phi_l(M_K,d)\left(\frac{t}{M_K^2}\right)^l\,.  
\end{equation}
Let us discard for a moment  the terms of order $t$ and higher 
in this expansion. It is then seen that
$F_{V,3}$ reduces to $F_{V,2}$, provided that we set
\begin{equation}
  \label{eq:l6L9}
l_6=-2L_9-\frac{1}{2}\Phi_0( M_K,d)\,.  
\end{equation}
At $d=4$, this relation  reduces to the one between the 
renormalised \lecs{} $l_6^\mathrm{r}$ and $L_9^\mathrm{r}$ worked out in Ref.~\cite{glnpb}, 
\begin{equation}
\label{eq:l6r}
l_6^\mathrm{r}(\mu)=
-2\,L_9^\mathrm{r}(\mu) + \frac{1}{192\pi^2}\left(\ln\frac{ B_0m_s}{\mu^2} + 1\right).
\end{equation}
This expressions is indeed of the form displayed in Eq.~(\ref{eq:matching}), with $d_{-1}=0$, whereas $d_0$ is simply the right hand side of Eq.~(\ref{eq:l6r}), generated by  the 
one--loop graphs considered here. 
We conclude that, at low energies, the expression of the vector form factor in
\chpt{3} reduces to the one in the two--flavour case, up to polynomial terms of
order $t^2$ and higher. An analogous
 statement holds true for  all Green functions of quark
currents built from up and down quarks alone, see below.

We  now come back to  the higher--order terms in Eq.~(\ref{eq:polynom}).
 We start with the observation that
the term of order $t^l$ contributes at order $t^{l+1}$ to $F_{V,2}$ -- those
with $l\geq 1$ are thus of the same chiral order in $F_{V,2}$ 
as the ones generated by graphs with $l+1$ loops in  \chpt{2}.
 Apparently, one runs into a problem with power counting here: the low--energy
 expansion of the one--loop contribution in \chpt{3}
amounts to terms of arbitrarily high orders in 
the $\mathrm{SU}(2)_R\times\mathrm{SU}(2)_L$ 
 expansion of $F_{V,2}$. Indeed, this is
 a rule rather than an exception: Because the strange quark mass is counted
 as a quantity of chiral order zero in  \chpt{2},
the counting of a quantity like
$t/M_K^2$ is different in the two theories.  As a result of this, 
higher--order loops in  \chpt{3} in general start to contribute already at 
leading order in \chpt{2}.
A  systematic and coherent scheme is  obtained by counting
 $n$--loop contributions -- and, in particular the 
relevant \lecs{} -- to be of order $\hbar^n$, and the strange quark mass to
 be of order $\hbar^{-1}$, see Refs.~\cite{LECs-p4,LECs-p6}.


\section{Matching of generating functionals in $\chi${\footnotesize
    PT}$\mathbf{_2}$ and $\chi${\footnotesize PT}$\mathbf{_3}$}
\label{sec:description}

We have developed in Ref.~\cite{LECs-p4} a generic method for the matching,
 which is based on the path 
integral formulation  of \chpt{}. The idea of this method is not to compare 
matrix elements that can be obtained in both formulations, 
but rather to restrict the three--flavour theory such that it only 
describes the same physics as the two--flavour formulation. 
Then, one compares their generating functionals containing 
all the Green functions and reads off the matching of the \lecs{}. 

The \lecs{}  do not depend on the light quark masses $m_u$ and $m_d$. 
Since both  theories are expansions around vanishing quark masses, 
we may set $m_u=m_d=0$ for the purpose of the matching.

The comparison of the generating functionals is in fact a comparison of all
possible Green functions, which depend on the external
fields. Obviously, they can only be compared with each other if they
depend on the \emph{same} external fields. Therefore, the external
fields of \chpt{3} need to be restricted to those of \chpt{2}. 
We also have to assure that the heavy mesons $K$ or $\eta$ running in
the loops do not have the possibility to go on--shell. Therefore, we
consider in addition the case where all external momenta are  small
 compared to
the kaon mass.  The physics of
$\mathrm{SU}(3)_R\times\mathrm{SU}(3)_L$ then reduces to the 
one of $\mathrm{SU}(2)_R\times\mathrm{SU}(2)_L$. We 
refer to this limit as the \emph{two--flavour limit}. 

The \lecs{} are the coefficients of local chiral operators in the
effective Lagrangian. Once one evaluates the generating functional
with the effective Lagrangian, besides the local terms also many non--local
contributions are generated, both in \chpt{2} as well as in \chpt{3}. 
However, the non--local contributions, appearing in \chpt{3}
as the result of low-energy expansion, will be exactly canceled by
\chpt{2} counterparts once the matching is performed. 
Therefore, to obtain the matching relations, it is sufficient to restrict oneself 
to the local parts in the evaluation of the generating functional of \chpt{3}.


\section{{\footnotesize LEC}s at order $\mathbf{p^2}$ and $\mathbf{p^4}$}
\label{sec:l2}

All the relations may be put in the form of Eq.~(\ref{eq:matching}).
To render the formulae more compact, we found it convenient to 
 slightly reorder the  expansions, such that they become 
a series in  the quantity $\mk^2$,
which stands for the one--loop expression of the (kaonmass)$^2$ in the limit
$m_u=m_d=0$, see e.g.~\cite{glnpb}.
The result is
\begin{eqnarray}
  \label{eq:1}
Y &=& Y_0\,\left[ 1 + a_Y\,x + b_Y\,x^2 + \cO(x^3) \right]
\co
\qquad
Y = F \, , \Sigma 
\co
\nn
l^\mathrm{r}_i &=& a_i + x\,b_i + \cO(x^2) \, , \quad (i\neq 7)\co
\qquad
l_7=\frac{F_0^2}{8B_0m_s}+a_7+x\,b_7+\cO(x^2)\co \\[1.5ex]
x &=& \frac{\mk^2}{N F_0^2} \, ,\qquad 
N = 16\pi^2 \, ,
\qquad
\Sigma = F^2 B \co
\qquad
\Sigma_0 = F_0^2 B_0 \fs
\nonumber
\end{eqnarray}
We denote the contributions proportional to $a_i\,\,\,(b_i)$ as \nlo{}\,\,(\nnlo{}) 
terms, generated by one--loop (two--loop) graphs in \chpt{3}.  Note that $l_7$ receives a contribution at leading
order (\lo{}) as well, proportional to $m_s^{-1}$, in agreement with the remarks made in the Introduction.
 The \lo{} and \nlo{} terms were  determined in Ref.~\cite{glnpb} more than 25 years ago, 
whereas the \nnlo{} terms  $b_i$ were only recently worked out \cite{LECs-p4}.
They  have the following structure,
\begin{equation}
  b = p_0 + p_1\,\ell_K + p_2\,\ell_K^2   
\co
\end{equation}
where $\ell_K = \ln(\bar{M}_K^2/\mu^2)$ is the chiral logarithm, and where we have dropped for simplicity  the index $i$.
The polynomials $p_j$ are independent of the strange quark mass, and 
their scale dependence is such that in combination with the logarithms 
it adds up to the scale independent quantity $b$.  In other words, the scale dependence of $l_i$ is exclusively generated by the one--loop contribution $a_i$.
The explicit results for the polynomials $p_j$  are
displayed in tables 2-4 of Ref.\cite{LECs-p4}. 

Let us  now illustrate,
in the case  of the low-energy constant $l_2^\mathrm{r}$,
 the strange quark mass dependence and the 
information one can obtain from the pertinent matching relation.
We found~\cite{LECs-p4} at two--loop order the result
\begin{eqnarray}
l_2^\mathrm{r} &=& 
- \frac{1}{24\,N}\,\left(\ell_K + 1 \right) + 4\,L^\mathrm{r}_{2}  
 +  x\,\,\Big\{ \frac{1}{N}\Big[
                                        \frac{433}{288}
                                      - \frac{1}{24}\ln\myfrac{4}{3}
                                      + \frac{1}{16}\,\rho_{1}
                                      \Big]
       -16\, N \,  \Big(2\, C^\mathrm{r}_{13}-C^\mathrm{r}_{11}\Big) 
\nonumber\\[2ex]
&&  
+ \, \Big[ \frac{13}{24\,N} - 8\,L^\mathrm{r}_{2}  - 2\,L_{3} \Big]
\, \ell_K  + \,  \frac{3}{8\,N}  \ell_K^2 
                                 \Big\}\co
\label{eq:l2}
\end{eqnarray}
where
\begin{eqnarray}
\rho_{1} &=& \sqrt{2}\,\mathrm{Cl}_2(\arccos(1/3)) 
\cong 1.41602
\co \qquad
\mathrm{Cl}_2(\theta) = -\frac{1}{2}\int_0^\theta\!\mathrm{d}\phi \,\, \ln\,\big(4\sin^2{\tfrac{\phi}{2}}\big)\fs
\label{eq:Clausen}
\end{eqnarray}
On the right--hand side, the $p^4$ ($p^6$) \lecs{} $L_{2,3}$ ($\Clr{11,13}$) occur, aside from 
known quantities.  Our definition of 
the $C_i$ differs from the one of Ref.~\cite{BCE} by a factor of $F^2_0$.
Explicitly, the $\L_6$--Lagrangian reads 
\begin{equation}
  \L_6^{\mathrm{SU}(2)} = F^{-2}\sum_{k=1}^{57}c_kP_k\co\qquad\L_6^{\mathrm{SU}(3)} = F_0^{-2}\sum_{k=1}^{94}C_kO_k
\end{equation}
for two and three flavours, respectively. [Note that the 57 
terms in $\L_6^{\mathrm{SU}(2)}$ are not independent \cite{Haefeli:2007ty}. We adhere to
the original notation used in Ref.~\cite{BCE} for later convenience.]

We now note that  $l_2^\mathrm{r}$ was determined in Ref.~\cite{Colangelo:2001df} 
from a dispersive analysis to rather high precision, and $L_{2,3}$ are also 
quite well known~\cite{Amoros:2001cp}. As a result of this, the relation (\ref{eq:l2}) allows one to 
 constrain the value of the  combination $\TClr$ \cite{LECs-p6}. We introduce the scale independent quantity
\begin{equation}
\bar l_2=48\pi^2 l_2^\mathrm{r}(\mu)-\ln{\frac{M_\pi^2}{\mu^2}},
\end{equation}
with $M_\pi$=139.57 MeV,
\begin{figure}[ht]
\centering
\begin{minipage}[]{1.0\linewidth}
\epsfig{file=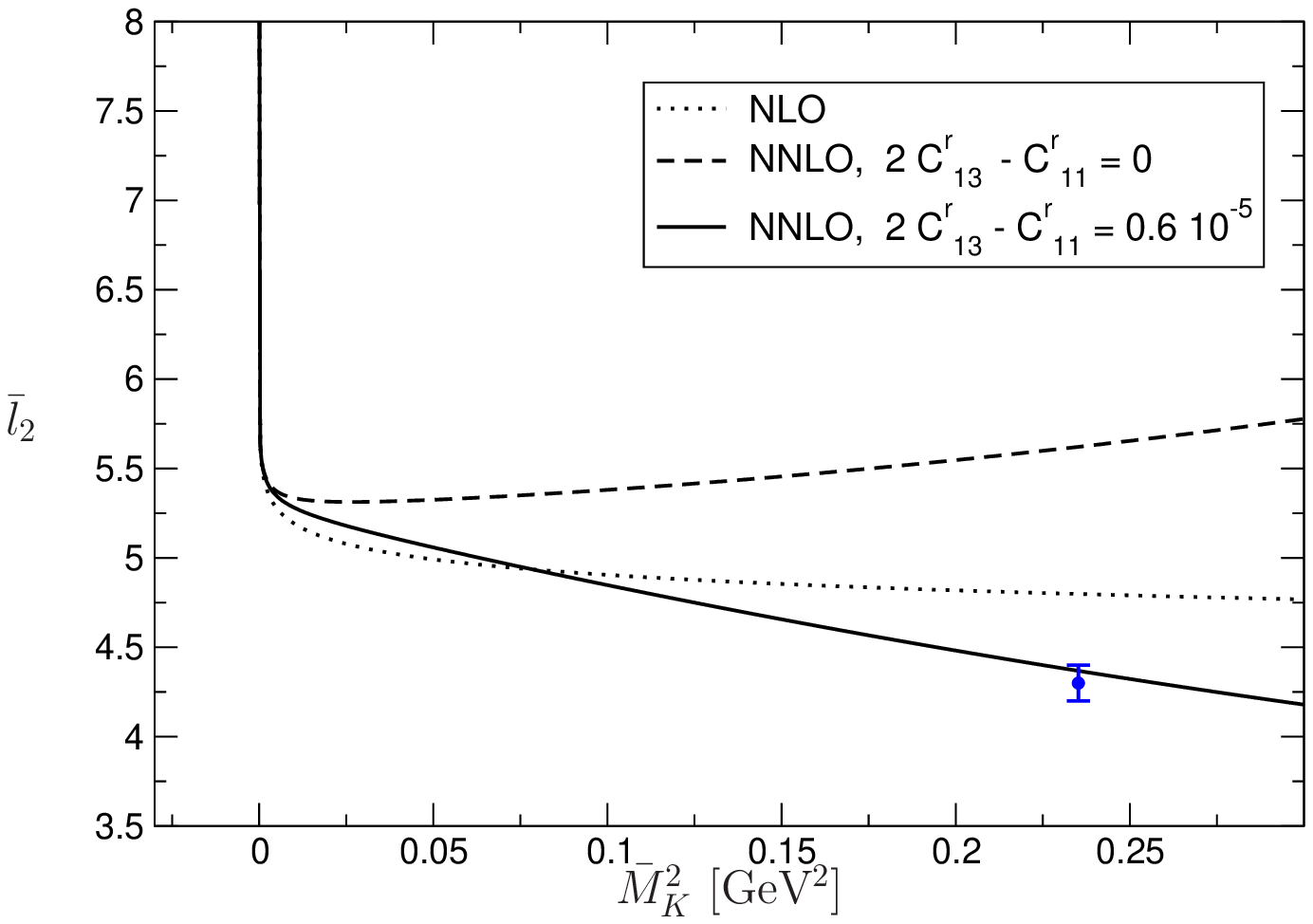,width=0.48\linewidth}
\hfill
\epsfig{file=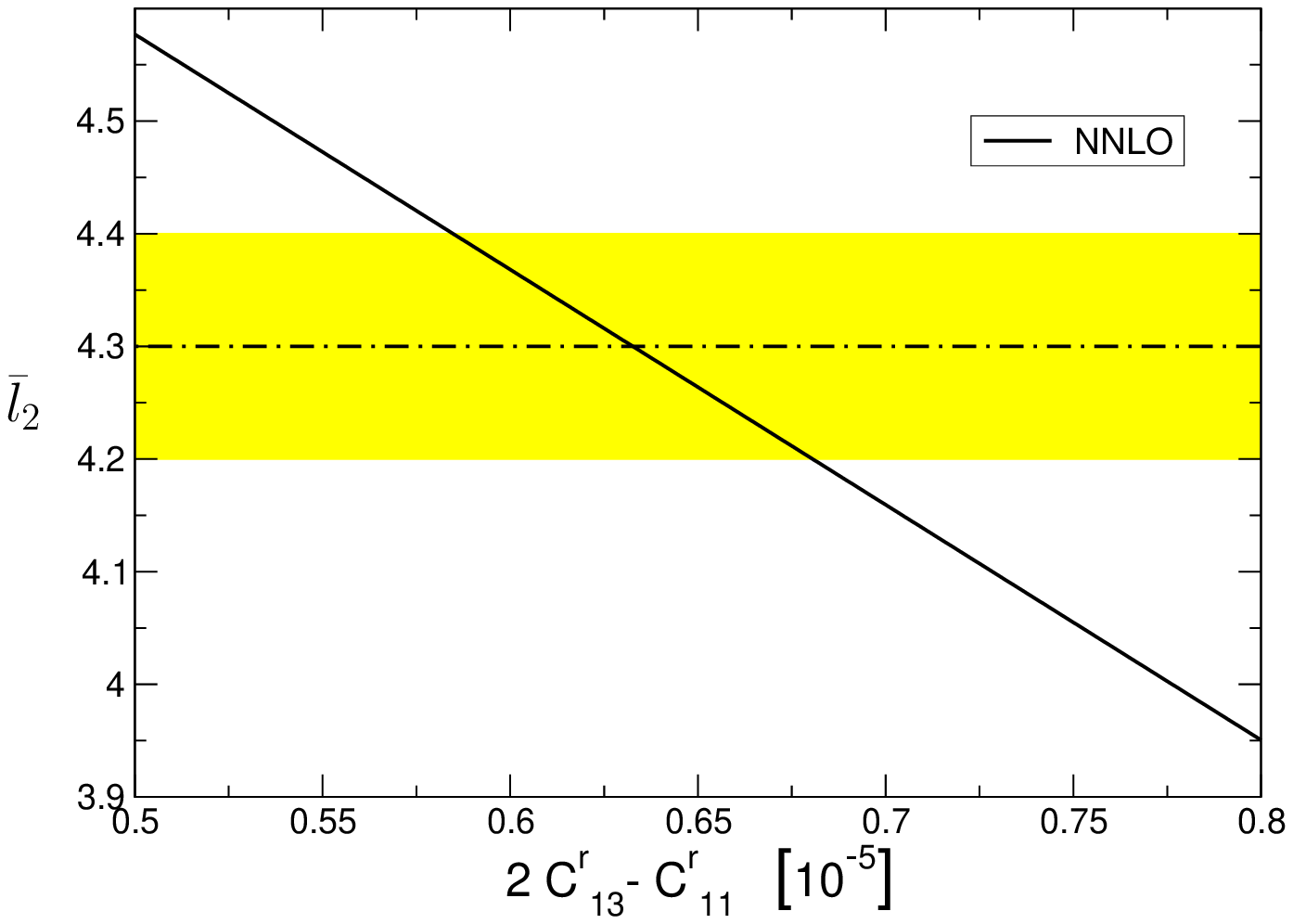,width=0.48\linewidth}\\
\end{minipage}
\caption{Left panel: Strange quark mass dependence of $\bar{l}_2$. As
  mentioned in the text, $\bar M_K$ denotes the
  kaon mass at one--loop accuracy in the limit
  $m_u=m_d=0$. The physical value of $m_s$ corresponds to 
  $\bar M_K\approx 485$ MeV.
  We show the \nlo{} (dotted line) as well as the \nnlo{} result with two
  choices for $\Clr{11,13}$: The dashed line 
  corresponds to $\TClr=0$, while the solid line is evaluated at 
  $\TClr=0.6\times 10^{-5}$,
  which reproduces the prediction from the dispersive analysis
  (data point with   small error bar). 
  Right panel: Dependence of $\bar{l}_2$ on the $p^6$ \lecs{}
  $\TClr$ at the physical value of $m_s$. The dashed--dotted line with
  the error band corresponds to the data point and its error bar in the
  left panel. The running scale is taken at
  $\mu=M_\rho=770$ MeV, and
$F_0=F_\pi = 92.4$ MeV.} 
\label{fig:l2.Mk2.eps}       
\end{figure}
 and illustrate the strange quark
mass dependence of $\bar{l}_2$  in Fig.\ref{fig:l2.Mk2.eps}
 (left panel), where $\bar l_2$ is shown as a function of $\mk^2$,
 at $\mu=M_\rho=770 \mathrm{\,MeV},
F_0=F_\pi = 92.4 \mathrm{\,MeV}$. 
The dotted line stands for the \nlo{} approximation, and
the \nnlo{} result is shown for two choices for  $\Clr{11,13}$:
the dashed (solid) line displays the 
case $\TClr = 0$ 
($\TClr = 0.6\cdot 10^{-5}$).
The solid line is constructed such that at the physical value of the strange quark 
mass, the \lec{} $\bar l_2$ agrees with the measured one  \cite{Colangelo:2001df} , 
within the uncertainties.

We shortly comment on the $\Clr{11,13}$  that occur in this application. In 
Ref.~\cite{Bijnens:2003xg}, $\Clr{13}$ is worked out from an 
analysis of scalar form factors. While the result is of the order 
$\TClr=0.6\cdot 10^{-5}$,
its precise value depends considerably on the input used, see table 2 in 
Ref.~\cite{Bijnens:2003xg} for more information.
In Ref.~\cite[table 12 (published version)]{ResCi} estimates for both \lecs{} $\Clr{11,13}$ are 
provided: the authors find that these  do not receive a contribution from
resonance exchange at leading order in large $N_C$ and therefore vanish 
at this order of accuracy. Because the scale at which this happens is not 
fixed a priori, that observation is not necessarily in contradiction 
with the above result.

The impact of these \lecs{}  on $\bar{l}_2$ is rather enhanced at physical
strange quark masses. This is illustrated in Fig.\ref{fig:l2.Mk2.eps}
(right panel). Taking the \lecs{} $L_j^\mathrm{r}$  at face value,
the window for a possible choice of the $\Clr{11,13}$ is then very narrow to be in
agreement with the data from a dispersive analysis.
To pin down the $\Clr{11,13}$ to good precision including an error analysis requires,
however, a more thorough exploration. In particular, one has  to take into
account that in the fits performed in Ref.~\cite{Amoros:2001cp}, 
an estimate of order $p^6$ counterterm contributions  was already used. 


\section{{\footnotesize LEC\rm}s at order $\mathbf{p^6}$}
\label{sec:results}

The evaluation of all matching relations at two--loop order for the \lecs{} at 
order $p^6$ is very complicated. To ease the calculations, we
did not deal with the full framework in  Ref.~\cite{LECs-p6},
but rather switched off the sources $s$ and $p$ (while retaining $m_s$). 
This yields the following simplifications:
\begin{enumerate}
\item[i)] the solution of the classical \eom{} 
for the eta--field is trivial, $\eta=0$;
\item[ii)] there is no  mixing between the $\eta$ and the $\pi^0$ fields.
\end{enumerate}
Point i) greatly simplifies the transition
from the \chpt{3} building blocks of the monomials to those of two flavours, as it
suppresses any effects from the eta, whereas point ii)
 eliminates many possible graphs and hence considerably reduces the 
requested labour. For example, in this restricted framework, 
the one--particle reducible graphs (two one--loop diagrams
linked by a single propagator) do not contribute to the
matching, see Ref.~\cite{LECs-p6}.

Aiming for the $\L_6$-monomials in the generating functional requires the evaluation of 
many graphs with sunset--like topology. In the two--flavour limit, where one is
interested in the local contributions only, one can simplify the
loop calculations by using a short distance expansion for the massive
propagators. This simplifies drastically the involved loop
integrals; however, the contributions from individual graphs are not chirally
invariant. Collecting terms stemming from different graphs to obtain a
manifestly chirally invariant result is rather cumbersome. Since we are
interested in the local terms only, we use a shortcut which is based 
on gauge invariance\footnote{We
are grateful to H.~Leutwyler for pointing out this possibility to us.}: 
 one may  choose a gauge such that  at some fixed space--time point $x_0$, 
the totally symmetric combination of up
to three derivatives acting on the chiral connection vanish,
\begin{equation}
  \label{eq:choose_gauge}
  \Gamma_\mu(x_0) = 0\co\d_{\{\mu}\Gamma_{\nu\}}(x_0) =
  0\co
  \d_{\{\mu}\d_{\nu}\Gamma_{\rho\}}(x_0) =
  0\co\d_{\{\mu}\d_{\nu}\d_\rho\Gamma_{\sigma\}}(x_0) = 0\fs
\end{equation}
Up to
four ordinary derivatives  are then indistinguishable
from the fully symmetric combinations of covariant derivatives: 
\begin{equation}
\d_{\mu}f(x_0) = \nabla_{\mu}f(x_0)\co
\d_\mu\d_\nu f(x_0) = \tfrac{1}{2}\{\d_\mu,\d_\nu\}f(x_0) = \tfrac12\{ \nabla_\mu,\nabla_\nu \}f(x_0)\co
\text{etc.} 
\end{equation}
This allows us to write even intermediate results in a manifestly chiral
invariant manner. 

To check our calculations in one corner, we matched the available \chpt{2}-- and
\chpt{3}--results for the vector--vector correlator \cite{Amoros:1999dp} 
 and for the pion form factor, 
worked out in Refs.~\cite{Bijnens:1998fm,Bijnens:2002hp}.
In this manner, we found that our relations for $\clr{56}$ and
$\clr{51}-\clr{53}$ 
agree with the results of Refs.~\cite{Bijnens:1998fm,Bijnens:2002hp}.  
Needless to say that this is quite a non--trivial check.

As already stated in Ref.~\cite{Haefeli:2007ty}, the monomial $P_{27}$ can be
discarded from the $p^6$--Lagrangian for \chpt{2}. Therefore, the matching
relations will certainly be a combination of some $\clr{i}$ and $\clr{27}$. 
Due to the restricted framework, only relations for \lecs{} not involving
monomials dependent on the sources $s$ or $p$ are nontrivial. In the
restricted framework, there is an additional relation among the remaining
$\SU{2}$--monomials:
\begin{equation}
\begin{split}
&
         \tfrac43P_{1}
       - \tfrac13P_{2}
       + P_{3}
       - \tfrac{10}{3}P_{24}
       + \tfrac43P_{25}
       + 2P_{26}
       - \tfrac83P_{28}
       - \tfrac12P_{29}
       + \tfrac12P_{30}
       - P_{31}
       + 2P_{32}
       - \tfrac12P_{33}
\\&
       + \tfrac43P_{36}
       - \tfrac43P_{37}
       - \tfrac{11}{6}P_{39}
       + \tfrac56P_{40}
       + \tfrac73P_{41}
       - \tfrac43P_{42}
       - \tfrac32P_{43}
       + \tfrac12P_{44}
       - \tfrac12P_{45}
       - P_{51}
       - P_{53}
=0\fs
\end{split}
\label{eq:new_relation}
\end{equation}
Because the \eom{} is different in the full framework, this relation is no 
longer valid there. We used Eq.~(\ref{eq:new_relation})  to exclude
the monomial $P_1$ from our consideration. As a result, we give
the matching for the \osn{27} combinations of $\clr{i}$ displayed in
table~\ref{tab:1}. In the full framework, an additional matching relation
(apart from the ones for the monomials involving the sources $s$ and $p$) for
$\clr{1}$ could be worked out, yielding the only missing piece in the matching for
the \osn{28} \lecs{} worked out here. 

The final result may be written  in the form
\begin{equation}
  \label{eq:p0p1p2}
  x_i = p^{(0)}_i + p^{(1)}_i\ell_K + p^{(2)}_i\ell_K^2+O(m_s)\co
\end{equation}
where $x_i$ denotes one of the \osn{27} linear combinations of the $\clr{i}$
displayed in table~\ref{tab:1}. The explicit expressions for
the polynomials $ p^{(n)}_i$ in the \chpt{3}--\lecs{} are displayed
in tables~2 and~3 of our article~\cite{LECs-p6}.

\renewcommand{\arraystretch}{1.5}
\newcommand{\extraline}{$\\ & $}
\setlength{\LTcapwidth}{\textwidth}
\begin{table}
\centering

\begin{tabular}{rl|rl|rl}

 $i$ &    $x_i$ & $i$ & $x_i$  & $i$ &  $x_i$   \\
\hline\hline
  1 & $ \clr{2}  + \frac{1}{4} \clr{1}$                            &  
 10 & $ \clr{32} - \frac{3}{2} \clr{1} - \clr{27}$                &
 19 & $ \clr{43} + \frac{9}{8} \clr{1} + \frac{1}{4} \clr{27}$      \\
%
  2 & $ \clr{3}  - \frac{3}{4} \clr{1}  $                         &
 11 & $ \clr{33} + \frac{3}{8} \clr{1} + \frac{1}{4} \clr{27}$     & 
 20 & $ \clr{44} - \frac{3}{8} \clr{1}- \frac{1}{4} \clr{27} $      \\
%
  3 & $ \clr{24} + \frac{5}{2} \clr{1}$                           &
 12 & $ \clr{36}- \clr{1}$                                         & 
 21 & $ \clr{45} + \frac{3}{8} \clr{1} + \frac{1}{4} \clr{27} $    \\
%
  4 & $ \clr{25} - \clr{1} $                                      &
 13 & $\clr{37} + \clr{1}$                                         & 
 22 & $ \clr{50} $                                                \\
%
  5 & $ \clr{26} - \frac{3}{2} \clr{1} $                           &
 14 & $ \clr{38}$                                                  & 
 23 & $ \clr{51} + \frac{3}{4} \clr{1}+ \frac{1}{2} \clr{27} $     \\
  6 & $ \clr{28}+ 2 \clr{1} - \clr{27} $                           &
 15 & $ \clr{39} + \frac{11}{8} \clr{1} + \frac{1}{4} \clr{27}$    &
 24 & $ \clr{52}$                                                 \\
%
  7 & $ \clr{29} + \frac{3}{8} \clr{1}  + \frac{1}{4} \clr{27} $    &
 16 & $ \clr{40} - \frac{5}{8} \clr{1} - \frac{1}{4} \clr{27} $     &
 25 & $ \clr{53}+ \frac{3}{4} \clr{1} + \frac{1}{2} \clr{27} $      \\
%
  8 & $ \clr{30} - \frac{3}{8} \clr{1} - \frac{1}{4} \clr{27}$    &
 17 & $ \clr{41} - \frac{7}{4} \clr{1} - \frac{1}{2} \clr{27}$    &
 26 & $ \clr{55}$                                                \\
%
  9 & $\clr{31}+ \frac{3}{4} \clr{1}  + \frac{1}{2} \clr{27}$      &
 18 & $ \clr{42} + \clr{1}$                                        &
 27 & $ \clr{56}$                                                 \\
\end{tabular}
\caption[]{\rule{0cm}{2ex}The quantities $x_i$ in Eq.~(\ref{eq:p0p1p2})}
\label{tab:1}
\end{table}


\section{Summary}
\label{sec:conclusion}

In this talk, we have discussed a general procedure~\cite{LECs-p4,LECs-p6} to work 
out the matching relations between 
the \lecs{} in \chpt{2} and \chpt{3} in a perturbative manner. 
 For the \lecs{} at order $p^2$ and  $p^4$, and for a subset 
of those at order $p^6$, the relations
 are now available at two--loop order.
The method could be used with only moderate adaption to work out more general
 matching relations, like the ones for \chpt{N-1} to \chpt{N}.
To obtain the matching relations of the 
the remaining  \lecs{}  at order $p^6$ to the same accuracy 
would require, however,  a very big amount of work.

We have  in addition illustrated the use of the results  in the case of
  $l_2^\mathrm{r}$: its precise knowledge, together
with the known values of $L_2^\mathrm{r}, L_3$, in principle allows one to
pin down the combination $\TClr$  rather precisely.

We refer the interested reader to our articles Refs.~\cite{LECs-p4,LECs-p6} 
for  the matching relations found,  
and for further details on the method used.


\ed